\documentclass[12pt, a4paper]{article}

\usepackage[utf8]{inputenc}
\usepackage{natbib}
\usepackage{authblk}
\usepackage{amsmath}
\usepackage{amssymb}
\usepackage{graphicx}
\usepackage{url}
\usepackage{booktabs}
\usepackage{float}
\usepackage[table]{xcolor}

\begin{document}

%--------------------------------------------------------------------

\title{Neural Speaker Diarization via Multilingual Training: Evaluation on Low-Resource Nepali-Hindi Speech}

%--------------------------------------------------------------------

\author[1]{Samip Neupane\thanks{077bct073.samip@pcampus.edu.np}}
\author[1]{Sandesh Pokhrel\thanks{077bct074.sandesh@pcampus.edu.np}}
\author[1]{Sandesh Pyakurel\thanks{077bct075.sandesh@pcampus.edu.np}}
\author[1]{Basanta Joshi\thanks{basanta@ioe.edu.np}}
\affil[1]{Department of Electronics and Computer Engineering, Pulchowk Campus, Institute of Engineering, Lalitpur, Nepal}
\date{}

%--------------------------------------------------------------------

\maketitle

\begin{abstract}
Speaker diarization, the task of determining "who spoke when" in a multi-speaker recording, is a critical component in applications such as meeting transcription, accessibility tools, and multilingual information retrieval. While end-to-end neural diarization systems have achieved strong performance for English and other high-resource languages, their effectiveness degrades substantially for underrepresented languages where annotated speech data is scarce.

This paper investigates speaker diarization for low-resource Nepali-Hindi speech through a multilingual training approach, comparing two modern architectures: EEND with encoder-decoder attractors (EEND-EDA) and EEND with Perceiver-based attractors (DiaPer). Both models are trained on a multilingual corpus combining English speech from LibriSpeech, diverse speaker recordings from VoxCeleb, and separately collected Nepali and Hindi audio, a setup designed to reduce language bias and encourage cross-lingual generalization. We evaluate both models across 2-speaker, 3-speaker, 4-speaker, and mixed-speaker scenarios on LibriSpeech, VoxCeleb, and Nepali-Hindi (NeHi) test sets. DiaPer achieves stronger overall performance than EEND-EDA, particularly in more challenging multi-speaker conditions, obtaining DERs of 3.28\%, 2.02\%, 4.05\%, and 4.76\% on NeHi 2-speaker, 3-speaker, 4-speaker, and mixed-speaker settings, respectively, compared to 1.50\%, 9.68\%, 16.17\%, and 11.19\% for EEND-EDA. These results demonstrate the viability of Perceiver-based end-to-end neural diarization for low-resource multilingual speech processing.

\vspace{1em}

\end{abstract}

%--------------------------------------------------------------------

\section{Introduction}
\label{introduction}
Automatic speaker diarization determines "who spoke when" in a multi-speaker recording, serving as a critical preprocessing step for applications such as meeting transcription, call-center analytics, and accessibility for the hearing impaired. Traditional diarization pipelines rely on feature extraction and clustering (e.g., x-vectors with Gaussian mixture models), which are not trained to optimize the diarization error rate directly and often fail in overlapping-speech scenarios. In contrast, recent end-to-end neural diarization (EEND) approaches model all speakers jointly as a multi-label sequence classification problem. For example, self-attentive EEND replaces recurrent layers with Transformer-based attention to capture global context across the entire recording \citep{fujita2019end}, thereby improving separation of overlapping speech segments.

Despite these advances in high-resource settings, there has been little work on underrepresented languages. This study focuses on Nepali and Hindi, two phonetically rich Indo-Aryan languages with limited diarization resources. Two EEND models are trained on a multilingual corpus: an encoder–decoder attractor network (EEND-EDA) \citep{horiguchi2020end} and a Perceiver-based attractor model (DiaPer) \citep{landini2024diaper}. The training corpus combines clean English speech from LibriSpeech, multilingual speaker recordings from VoxCeleb spanning European, Indo-Aryan, and other language communities, along with separately collected Nepali and Hindi recordings. Multi-speaker training samples are synthesized by merging isolated speaker recordings from these sources. Evaluation is conducted on LibriSpeech, VoxCeleb, and Nepali-Hindi (NeHi) test sets, and findings suggest that DiaPer provides consistent improvements over EEND-EDA in terms of Diarization Error Rate (DER), particularly on low-resource Nepali-Hindi data.

%--------------------------------------------------------------------

\section{Related Work}
\label{related_work}
Early speaker diarization systems relied on segment-level speaker embeddings and clustering techniques. These approaches, such as i-vector and x-vector based pipelines with agglomerative clustering, are not jointly optimized and struggle in overlapping speech scenarios. End-to-end neural diarization (EEND) methods aim to overcome these limitations by learning speaker labels directly from the audio input.

Fujita \textit{et al.} \citep{fujita2019end} proposed a self-attention-based EEND model (SA-EEND) that replaces the LSTM layers in previous EEND models with Transformer-based self-attention. This modification improves the model’s capacity to capture long-range dependencies and has demonstrated superior performance on overlapping speech tasks.

Horiguchi \textit{et al.} \citep{horiguchi2020end} extended this framework to handle an unknown number of speakers by introducing encoder–decoder attractors (EEND-EDA). This architecture computes a dynamic number of attractors using an LSTM decoder, which are then used to predict per-speaker activity. EEND-EDA outperformed fixed-speaker-count baselines on datasets like CALLHOME, where speaker count varies across recordings.

More recently, Landini \textit{et al.} \citep{landini2024diaper} proposed DiaPer, an EEND architecture that replaces the recurrent decoder with a Perceiver-based attractor module. The Perceiver model uses cross-attention to compress variable-length inputs into a fixed set of latent vectors, providing scalability and faster inference. DiaPer achieves lower DERs than EEND-EDA on several datasets, including CallHome, while offering better generalization and computational efficiency.

These EEND-based approaches motivate the current investigation into applying DiaPer and EEND-EDA for low-resource language diarization, particularly for Nepali-Hindi.

%--------------------------------------------------------------------

\section{The Models}
\label{the_models}
Two end-to-end neural diarization (EEND) models were used with encoder-decoder-based attractors and perceiver-based attractors each for the comparative study.

\subsection{EEND-EDA: Encoder-Decoder Attractor Model}
EEND-EDA improves over the original EEND framework by incorporating an encoder-decoder attractor mechanism which handles a larger number of speakers, unlike the latter having limited output size in its architecture \citep{horiguchi2020end}.

The attractor vectors, each designed to represent a unique speaker in the segment, are matched with the encoder output to compute the probability that each speaker is active at a given time step. The interaction between attractors and embeddings is captured through a dot-product operation followed by a sigmoid activation:
\begin{equation}
\hat{Y} = \sigma(A^\top E), \label{eq:posterior}
\end{equation}
where $E \in \mathbb{R}^{D \times T}$ represents the encoder output and $A \in \mathbb{R}^{D \times S}$ represents the attractor vectors. The result $\hat{Y} \in (0, 1)^{S \times T}$ provides the probability of speaker activity in each time frame. This equation follows the approach introduced in  \citep{horiguchi2020end}.

\subsection{EEND-DiaPer: Perceiver-Based Attractor Model}
The EEND-DiaPer uses cross-attention to project variable input to a fixed set of compact latent representations from which attractors are derived \citep{landini2024diaper}.

\begin{equation}
\hat{Y}^{(l-1)} = \sigma\left(E^{(l-1)} \cdot \text{PercDec}(E^{(l-1)})^\top\right), \label{eq:diaper_conditioning}
\end{equation}
where $\text{PercDec}(\cdot)$ denotes the Perceiver-based attractor decoder, and $\sigma(\cdot)$ is the sigmoid activation function \citep{landini2024diaper}. This mechanism enables the frame-level embeddings to be modulated based on the inferred attractor context, improving temporal coherence and speaker separation.

To enhance speech overlap handling, Landini et al. \citep{landini2024diaper} introduce auxiliary losses computed using intermediate attractors after each Perceiver block. Specifically, rather than relying solely on the final diarization and attractor existence losses, DiaPer applies block-wise supervision at every layer of the Perceiver decoder using the frame embeddings produced by the frame encoder. This design enforces progressively better attractor quality throughout the network depth, an approach not used natively in EEND-EDA.

In summary, both models leverage end-to-end architectures with permutation-invariant training, but differ in how speaker characteristics are captured and leveraged: EEND-EDA uses encoder-decoder attractors through LSTM decoder, while EEND-DiaPer uses a perceiver-based latent mechanism through perceiver-based decoder.

%--------------------------------------------------------------------

\section{Experimental Setup}
\label{experimental_setup}
\subsection{Data Collection}
A multilingual dataset was compiled from four sources:
\begin{itemize}
    \item LibriSpeech \citep{panayotov2021librispeech}: 921 English speakers (clean studio recordings).
    \item VoxCeleb \citep{nagrani2017voxceleb}: 1,211 speakers with real-world variability.
    \item Nepali Female Speakers \citep{kjartansson-etal-tts-sltu2018}: 18 native Nepali speakers.
    \item Hindi Audio \citep{5vgy-yb08-20}: 100 speakers for South Asian linguistic diversity.
\end{itemize}

\subsection{Data Preparation}
\textbf{Voice Activity Detection (VAD)}: WebRTC Voice Activity Detection \citep{pywebrtcvad} library was employed to isolate speech segments from raw audio. The VAD was configured to its most aggressive mode (mode 3), prioritizing speech retention in noisy or mixed environments. Audio was segmented into 30 ms frames, and only frames positively classified as speech were retained.

\textbf{Multi-Speaker Synthesis}: Multi‑speaker recordings were generated by merging single‑speaker clips using the Python pydub AudioSegment library \citep{robert2018pydub}. Each clip was first resampled and normalized to a uniform 16 kHz, 16‑bit PCM format, then concatenated with configurable silence intervals to emulate natural speaker turns.

\begin{itemize}
    \item \textbf{Merge Parameters}:
    \begin{itemize}
        \item Speaker count: 2/3/4 per clip (min/max speakers).
        \item Utterances per speaker: 1–3 (randomized).
        \item Silence insertion: 50\% probability, 500–2000 ms gaps (randomized).
    \end{itemize}
\end{itemize}

\textbf{Merged Dataset Summary}:
\begin{itemize}
    \item \textbf{Kaldi Metadata}: Utterance-speaker mappings and timing data were stored in Kaldi-format files:
    
    \texttt{wav.scp}: Audio file paths.
    
    \texttt{utt2spk}: Utterance-to-speaker mappings.
    
    \texttt{segments}: Start/end times of speech segments.
    
    \texttt{reco2dur}: Recording durations.
\end{itemize}

The dataset is partitioned into train, validation, and test subsets with a unique speaker count ratio of 4:1:1, comprising a total of 2,150 unique speakers drawn from four distinct data sources.

The distribution of merged audio files used for training, and validation is shown in Table~\ref{tab:merged-audio}.
\begin{table}[ht]
\centering
\caption{Summary of Merged Multi-Speaker Audio Dataset}
\label{tab:merged-audio}
\begin{tabular}{|l|p{3cm}|c|}
\hline
\textbf{Speaker Type} & \textbf{Data Source} & \textbf{Sample Count} \\
\hline
2 speakers & Merged single-speaker datasets & 
\begin{tabular}[c]{@{}c@{}}Train: 10000\\ Val: 3000\end{tabular} \\
\hline
3 speakers & Merged single-speaker datasets & 
\begin{tabular}[c]{@{}c@{}}Train: 10000\\ Val: 3000\end{tabular} \\
\hline
4 speakers & Merged single-speaker datasets & 
\begin{tabular}[c]{@{}c@{}}Train: 10000\\ Val: 3000\end{tabular} \\
\hline
2,3,4 mix & Merged single-speaker datasets & 
\begin{tabular}[c]{@{}c@{}}Train: 10000\\ Val: 3000\end{tabular} \\
\hline
\end{tabular}
\end{table}

\subsection{Feature Extraction \& Normalization}
40-dimensional log Mel spectrograms were computed using 25ms Hamming windows with 10ms stride, derived from the raw audio sampled at 16 kHz. The features were normalized to zero mean to mitigate variability across recording environments. This preprocessing was consistent for all models, ensuring spectral representations aligned with the input requirements of both the perceiver and transformer architectures.

\subsection{Model Training}
Two models were trained:
 
\textbf{EEND-EDA} \citep{horiguchi2020end}: LSTM-based encoder-decoder to iteratively generate speaker attractors.

\textbf{DiaPer} \citep{landini2024diaper}:  Replaces LSTM with Perceiver-based cross attention, aggregating frame embeddings into fixed-size latent vectors.

Both diarization architectures - \textbf{DiaPer} and \textbf{EEND‑EDA} were trained using identical configurations. Each model was trained on 40-dimensional acoustic features with a frame size of 400 samples. Training was performed using the Adam optimizer with a learning rate of $1 \times 10^{-5}$, a batch size of 32 for training, and 128 for validation. This consistent setup ensures a fair comparison between the two architectures under equivalent training conditions.

During training, different learning rates were tested. Lower values, such as \(10^{-6}\), resulted in very slow convergence and required many epochs, while higher values led to instability or suboptimal performance. A learning rate of \(10^{-5}\) was found to provide the best balance between stability and efficiency and was therefore chosen for the experiments.

\subsection{Training Infrastructure Constraints}
Due to computational constraints, each speaker scenario (2/3/4/mixed) was trained sequentially, with each phase initialized from the checkpoint of the previous phase. EEND-EDA was trained for 40 epochs per phase (40/40/40/40 = 160 total epochs), while DiaPer was trained for 50/50/45/30 epochs per phase (175 total epochs), with the reduced epochs in later phases reflecting earlier convergence. All experiments were conducted on a single GPU (NVIDIA T4, 16GB VRAM).

\subsection{Dataset Composition per Phase}

\begin{table}[H]
  \centering
  \footnotesize
  \caption{Training Sets per Phase}
  \renewcommand{\arraystretch}{1.1}

  \begin{tabular}{@{}lcccc@{}}
     \toprule
     \textbf{Phase} & \textbf{Train Audio Count}   & \textbf{Train Audio Length (hr)} \\
     \midrule
    2‑speaker      & 10000                       & 145.10                       \\
    3‑speaker      & 10000                       & 161.37                      \\
    4‑speaker      & 10000                       & 215.90                      \\
    M‑speaker      & 10000                       & 171.12                      \\
    \bottomrule
  \end{tabular}
\end{table}

\begin{table}[H]
  \centering
  \footnotesize
  \caption{Validation Sets per Phase}
  \renewcommand{\arraystretch}{1.1}

  \begin{tabular}{@{}lcccc@{}}
     \toprule
     \textbf{Phase} &\textbf{Validation Audio Count}   & \textbf{Val Audio Length (hr)} \\
     \midrule
    2‑speaker      & 3000                       & 37.67                      \\
    3‑speaker      & 3000                       & 48.49                      \\
    4‑speaker      & 3000                       & 64.53                      \\
    M‑speaker      & 3000                       & 59.60                       \\
    \bottomrule
  \end{tabular}
\end{table}

\subsection{Training Procedure}

We trained both DiaPer and EEND‑EDA models on simulated audio mixtures containing 2 to 4 speakers, as well as on datasets with mixed speaker counts. Each model was trained using 10,000 audio clips and validated on 3,000 samples. Throughout training, we monitored the loss functions and diarization error rate (DER), including its components: missed speech (DER\_miss) and false alarm (DER\_FA).

\paragraph{DiaPer}  
DiaPer employs a frame-level activation loss optimized using a perceiver-based attractor mechanism. During training, both the activation loss and DER consistently decreased on the training set, with validation metrics closely following. This trend persisted across scenarios with 2, 3, and 4 speakers, as well as in mixed speaker count conditions, indicating effective learning and good generalization with minimal overfitting.

\paragraph{EEND‑EDA}  
EEND‑EDA utilizes an end-to-end loss that jointly penalizes missed speech and false alarms through a permutation-free objective. In the 2-speaker scenario, both loss and DER\_miss decreased steadily, and DER\_FA converged. However, as the number of speakers increased to 3 and 4, DER\_FA began to fluctuate and showed a gradual increase, despite continued decreases in loss and DER\_miss. This suggests that while the model learns effectively, it becomes more prone to false alarms with higher speaker counts. The same pattern was observed in mixed speaker count training.

\paragraph{Summary}  
Overall, both models demonstrated effective learning with decreasing loss and DER metrics. DiaPer maintained stable performance across varying speaker counts, showing strong generalization and minimal overfitting. In contrast, EEND‑EDA exhibited increased sensitivity to false alarms as the number of speakers increased, indicating challenges in handling more complex speaker scenarios.

%--------------------------------------------------------------------

\section{Results}
\label{results}
\subsection{Diarization Error Rate (DER) Calculation}
The Diarization Error Rate (DER) is computed as follows:

\begin{equation}
DER = \frac{\text{Missed Speech} + \text{False Alarms} + \text{Confusion Errors}}{\text{Total Speech Time}}
\end{equation}

Where:
\begin{itemize}
    \item \textbf{Missed Speech}: Total duration of unidentified speech
    \item \textbf{False Alarms}: Total duration of incorrectly identified speech
    \item \textbf{Confusion Errors}: Total duration of speaker label mismatches during overlap
    \item \textbf{Total Speech Time}: Sum duration of all ground truth speech segments
\end{itemize}

DER was calculated using PyAnnote's DiarizationErrorRate \citep{pyannote.metrics} by comparing reference and hypothesis segments. During evaluation, a 0.25-second collar was applied at segment boundaries, following standard speaker diarization evaluation protocols.

\subsection{Testing Datasets}
Performance was evaluated on three benchmark datasets with varying speaker counts:
\begin{itemize}
    \item \textbf{LibriSpeech} (2/3/4/Mspk)
    \item \textbf{NeHi} (2/3/4/Mspk) (Nepali and Hindi)
    \item \textbf{VoxCeleb} (2/3/4/Mspk)
\end{itemize}

\begin{table}[H]
    \centering
    \caption{Testing Dataset Composition}
    \label{tab:dataset_stats}
    \footnotesize
    \begin{tabular}{lcc}
        \toprule
        \textbf{Dataset} & \textbf{Audio Samples} & \textbf{Duration (hr)} \\
        \midrule
        LibriSpeech-2spk & 1,000 & 11.87 \\
        LibriSpeech-3spk & 1,000 & 17.95 \\
        LibriSpeech-4spk & 1,000 & 24.21 \\
        LibriSpeech-Mspk & 1,000 & 18.52 \\
        NeHi-2spk & 100 & 0.76 \\
        NeHi-3spk & 100 & 1.19 \\
        NeHi-4spk & 100 & 1.64 \\
        NeHi-Mspk & 100 & 1.21 \\
        VoxCeleb-2spk & 2,000 & 19.66 \\
        VoxCeleb-3spk & 2,000 & 29.91 \\
        VoxCeleb-4spk & 2,000 & 39.84 \\
        VoxCeleb-Mspk & 2,000 & 29.46 \\
        \bottomrule
    \end{tabular}
\end{table}

\subsection{Diarization Performance}
We evaluated four model variants of both \textbf{DiaPer}\citep{landini2024diaper} and \textbf{EEND-EDA}\citep{horiguchi2020end}, each trained for specific speaker scenarios:
\begin{itemize}
\item 2-speaker (2spk)
\item 3-speaker (3spk)
\item 4-speaker (4spk)
\item Mixed-speaker (Mspk)
\end{itemize}

\begin{table}[H]
    \centering
    \caption{DER (\%) on LibriSpeech Test Sets}
    \label{tab:librispeech_results}
    \footnotesize
    \begin{tabular}{l*{4}{p{0.5cm}}|*{4}{p{0.5cm}}}
        \toprule
        \textbf{Test Sets} & \multicolumn{4}{c}{\textbf{DiaPer}} & \multicolumn{4}{c}{\textbf{EEND}} \\
        \cmidrule(lr){2-5} \cmidrule(lr){6-9}
        & 2spk & 3spk & 4spk & Mspk & 2spk & 3spk & 4spk & Mspk \\
        \midrule
        2-speaker & \cellcolor{gray!30}1.55 & 8.25 & 7.53 & \cellcolor{brown!30}5.22 & \cellcolor{gray!30}1.43 & 5.21 & 6.21 & \cellcolor{brown!30}4.89 \\
        3-speaker & 23.32 & \cellcolor{gray!30}2.99 & 5.59 & \cellcolor{brown!30}5.29 & 37.16 & \cellcolor{gray!30}8.31 & 10.68 & \cellcolor{brown!30}8.95 \\
        4-speaker & 37.17 & 20.24 & \cellcolor{gray!30}5.56 & \cellcolor{brown!30}7.71 & 56.53 & 25.50 & \cellcolor{gray!30}18.73 & \cellcolor{brown!30}18.41 \\
        Mixed & 21.45 & 9.67 & 5.66 & \cellcolor{gray!60}5.70 & 43.63 & 18.15 & 11.48 & \cellcolor{gray!60}10.50 \\
        \bottomrule
    \end{tabular}
\end{table}

\begin{table}[H]
    \centering
    \caption{DER (\%) on NeHi Test Sets}
    \label{tab:nehi_results}
    \footnotesize
    \begin{tabular}{l*{4}{p{0.5cm}}|*{4}{p{0.5cm}}}
        \toprule
        \textbf{Test Sets} & \multicolumn{4}{c}{\textbf{DiaPer}} & \multicolumn{4}{c}{\textbf{EEND}} \\
        \cmidrule(lr){2-5} \cmidrule(lr){6-9}
        & 2spk & 3spk & 4spk & Mspk & 2spk & 3spk & 4spk & Mspk \\
        \midrule
        2-speaker & \cellcolor{gray!30}3.28 & 5.31 & 2.99 & \cellcolor{brown!30}0.88 & \cellcolor{gray!30}1.50 & 4.00 & 4.73 & \cellcolor{brown!30}3.19 \\
        3-speaker & 24.44 & \cellcolor{gray!30}2.02 & 2.51 & \cellcolor{brown!30}4.16 & 44.28 & \cellcolor{gray!30}9.68 & 10.80 & \cellcolor{brown!30}8.96 \\
        4-speaker & 35.34 & 18.90 & \cellcolor{gray!30}4.05 & \cellcolor{brown!30}8.20 & 68.14 & 25.94 & \cellcolor{gray!30}16.17 & \cellcolor{brown!30}15.44 \\
        Mixed & 21.73 & 9.02 & 5.83 & \cellcolor{gray!60}4.76 & 55.50 & 19.07 & 10.64 & \cellcolor{gray!60}11.19 \\
        \bottomrule
    \end{tabular}
\end{table}

\begin{table}[H]
    \centering
    \caption{DER (\%) on VoxCeleb Test Sets}
    \label{tab:voxceleb_results}
    \footnotesize
    \begin{tabular}{l*{4}{p{0.5cm}}|*{4}{p{0.5cm}}}
        \toprule
        \textbf{Test Sets} & \multicolumn{4}{c}{\textbf{DiaPer}} & \multicolumn{4}{c}{\textbf{EEND}} \\
        \cmidrule(lr){2-5} \cmidrule(lr){6-9}
        & 2spk & 3spk & 4spk & Mspk & 2spk & 3spk & 4spk & Mspk \\
        \midrule
        2-speaker & \cellcolor{gray!30}1.14 & 5.06 & 5.78 & \cellcolor{brown!30}2.44 & \cellcolor{gray!30}1.82 & 4.41 & 5.61 & \cellcolor{brown!30}3.09 \\
        3-speaker & 21.20 & \cellcolor{gray!30}1.11 & 1.89 & \cellcolor{brown!30}1.73 & 42.71 & \cellcolor{gray!30}7.06 & 8.41 & \cellcolor{brown!30}6.95 \\
        4-speaker & 33.93 & 17.08 & \cellcolor{gray!30}1.94 & \cellcolor{brown!30}3.17 & 65.80 & 23.33 & \cellcolor{gray!30}17.40 & \cellcolor{brown!30}16.91 \\
        Mixed & 18.48 & 7.04 & 2.87 & \cellcolor{gray!60}2.60 & 53.14 & 18.40 & 10.01 & \cellcolor{gray!60}8.99 \\
        \bottomrule
    \end{tabular}
\end{table}

\subsection{Analysis of Diarization Performance}
The results demonstrate that DiaPer outperformed EEND-EDA across most conditions, particularly in challenging multilingual (NeHi) and variable speaker scenarios. For same-speaker configurations (diagonal gray cells), DiaPer achieves significantly lower DERs than EEND-EDA, with NeHi-3spk showing an absolute improvement (2.02\% vs 9.68\%). This advantage amplifies with increasing speaker counts: DiaPer's Mspk model achieves 4.76\% DER on NeHi-Mspk compared to EEND-EDA's 11.19\%, indicating superior adaptation to low-resource multilingual data. 

On the LibriSpeech dataset, the mixed-speaker model of DiaPer achieves 5.70\% DER compared to EEND-EDA's 10.50\%, suggesting better generalization across 
varying speaker counts.

DiaPer’s Mspk model achieves 2.60\% DER on VoxCeleb-Mspk, significantly outperforming EEND-EDA’s 8.99\%. These results suggest DiaPer’s architectural innovations better handle cross-lingual variations while maintaining flexibility across speaker counts.

The only exception is the 2-speaker condition on NeHi and LibriSpeech test sets, where EEND-EDA achieves marginally lower DERs (1.50\% vs 3.28\% and 1.43\% vs 1.55\% respectively), suggesting that the simpler attractor mechanism of EEND-EDA may be sufficient for lower-complexity scenarios.
%--------------------------------------------------------------------

\section{Conclusion}
\label{conclusion}
This study compared two end-to-end neural diarization architectures, DiaPer and EEND-EDA, trained on a multilingual corpus and evaluated on low-resource Nepali-Hindi speech. DiaPer outperforms EEND-EDA across most multi-speaker conditions, particularly in 3-speaker, 4-speaker, and mixed-speaker scenarios, demonstrating the advantage of Perceiver-based attractors for cross-lingual generalization. However, EEND-EDA shows marginally better performance in the 2-speaker condition, suggesting that architectural trade-offs remain scenario-dependent.

Several limitations should be acknowledged. The Nepali and Hindi test data are evaluated as a combined set, which prevents language-specific error analysis. Training data for Nepali is limited to 18 female speakers, introducing potential speaker diversity bias. Furthermore, all evaluation is conducted on synthetically generated multi-speaker recordings rather than real conversational audio, which may not fully reflect real-world diarization challenges.

The most meaningful next step would be training and evaluation on real conversational recordings with human-annotated diarization labels. For low-resource languages like Nepali and Hindi, such data remains scarce and costly to produce, but is essential for validating whether these models generalize beyond synthetic conditions to practical deployment scenarios.
%--------------------------------------------------------------------

\bibliographystyle{unsrt}
\bibliography{references}

\end{document}